\documentclass{aastex63}
\usepackage{CJK}

\usepackage{newtxtext,newtxmath}
\usepackage[T1]{fontenc}
\usepackage{ae,aecompl}
\usepackage{graphicx}	
\usepackage{amsmath}	
\usepackage{amssymb}	

\newcommand{\sigcrb}{\mbox{$\sigma^{2}$ CrB}}

\newcommand{\vsini}{\mbox{$v\,\sin\,i$}}

\received{}
\revised{}
\accepted{}
\submitjournal{ApJ}

\shorttitle{Doppler imaging of \sigcrb}
\shortauthors{Xiang et al.}

\begin{document}
\begin{CJK*}{UTF8}{gbsn}
\title[Doppler imaging of \sigcrb]{Doppler imaging and differential rotation of $\sigma^{2}$ Coronae Borealis using SONG\footnote{Based on observations made with the Hertzsprung SONG telescope operated on the Spanish Observatorio del Teide on the island of Tenerife by the Aarhus and Copenhagen Universities and by the Instituto de Astrof\'{i}sica de Canarias.}}

\correspondingauthor{Shenghong Gu, Yue Xiang}
\email{shenghonggu@ynao.ac.cn, xy@ynao.ac.cn}

\author{Yue Xiang}
\affiliation{Yunnan Observatories, Chinese Academy of Sciences, Kunming 650216, China}
\affiliation{Key Laboratory for the Structure and Evolution of Celestial Objects, Chinese Academy of Sciences, Kunming 650216, China}

\author{Shenghong Gu}
\affiliation{Yunnan Observatories, Chinese Academy of Sciences, Kunming 650216, China}
\affiliation{Key Laboratory for the Structure and Evolution of Celestial Objects, Chinese Academy of Sciences, Kunming 650216, China}
\affiliation{School of Astronomy and Space Sciences, University of Chinese Academy of Sciences, Beijing 101408, China}

\author{A. Collier Cameron}
\affiliation{School of Physics and Astronomy, University of St Andrews, Fife KY16 9SS, UK}

\author{J. R. Barnes}
\affiliation{Department of Physical Sciences, The Open University, Walton Hall, Milton Keynes MK7 6AA, UK}

\author{J. Christensen-Dalsgaard}
\affiliation{Stellar Astrophysics Centre (SAC), Department of Physics and Astronomy, Aarhus University, Ny Munkegade 120, DK-8000 Aarhus, Denmark}

\author{F. Grundahl}
\affiliation{Stellar Astrophysics Centre (SAC), Department of Physics and Astronomy, Aarhus University, Ny Munkegade 120, DK-8000 Aarhus, Denmark}

\author{V. Antoci}
\affiliation{Stellar Astrophysics Centre (SAC), Department of Physics and Astronomy, Aarhus University, Ny Munkegade 120, DK-8000 Aarhus, Denmark}

\author{M. F. Andersen}
\affiliation{Stellar Astrophysics Centre (SAC), Department of Physics and Astronomy, Aarhus University, Ny Munkegade 120, DK-8000 Aarhus, Denmark}

\author{P. L. Pall\'{e}}
\affiliation{Instituto de Astrof\'{i}sica de Canarias, E-38205 La Laguna, Tenerife, Spain}
\affiliation{Departamento de Astrof\'{i}sica, Universidad de La Laguna, E-38205 La Laguna, Tenerife, Spain}

\begin{abstract}
We present new Doppler images of both components of the double-lined binary \sigcrb, based on the high-resolution spectroscopic data collected during 11 nights in 2015 March--April. The observed spectra form two independent data sets with sufficient phase coverage. We apply the least-squares deconvolution to all observed spectra to obtain high signal-to-noise mean profiles, from which we derive the Doppler images of both components of \sigcrb\ simultaneously. The surfaces of both F9 and G0 components are dominated by pronounced polar spots. The F9 component exhibits a weak spot at latitude 30\degr\ and its mid-to-low latitudes are relatively featureless. The G0 star shows an extended spot structure at latitude 30\degr, and its surface spot coverage is larger than that of the F9 star, which suggests a higher level of magnetic activity. With the cross-correlation method, we derive a solar-like surface differential rotation on the G0 star of \sigcrb\ for the first time, and the surface shear rate is $\Delta \Omega = 0.180 \pm 0.004$ rad d$^{-1}$ and $\alpha = \Delta \Omega / \Omega_{eq} = 0.032 \pm 0.001$. We do not obtain a clear surface shear law for the F9 star due to the lack of mid-to-low latitude features, but detect a systematic longitude shift of high-latitude spots, which indicates a slower rotation with respect to the co-rotating frame.
\end{abstract}


\section{Introduction}
Solar-like stars with a convection envelope show cool starspots caused by strong local magnetic field on stellar surfaces. Doppler imaging is a powerful technique for the study of spot activity on rapidly rotating stars. Recently, various active single and binary stars have been investigated by means of the Doppler imaging technique \citep{str2009}. Unlike that on the Sun, persistent large high-latitude or polar spots are detected on many active stars with different stellar parameters (e.g. \citealt{cam1994,rice2001,hac2019}). The magnetic flux is affected by dominant Coriolis force within the convection zone of rapid rotators to form high-latitude magnetic fields \citep{sch1996,gra2000}. The rising magnetic flux-tube can also be affected by the tidal force to emerge at preferred longitudes \citep{hol2003}, which are present on many active binary systems (e.g. \citealt{ber1998}, \citealt{olah2006}).

Differential rotation plays an important role in the stellar dynamo. Up to now, the solar-like latitude-dependent surface shear is detected on both of single and binary stars, with the help of the Doppler imaging (e.g. \citealt{bar2000,dun2008,kri2014,ozd2016}). Some stars show anti-solar differential rotations (e.g. \citealt{web2007,kov2007,kov2017,har2016}), which means the equator rotates slower than the stellar pole. \citet{gas2014} suggested that the direction of differential rotation is determined by the Rossby number and the domination of the Coriolis force.

\sigcrb\ (TZ CrB, HD 146361, HR 6063) is a component of the multiple system $\sigma$ Coronae Borealis. Other components include a solar-like star $\sigma^{1}$ CrB and an M dwarf binary \citep{rag2009}. \sigcrb\ is an active, double-lined binary, consisting of an F9 and a G0 stars, with an orbital period of about 1.14 d and an inclination degree of 28\degr\ \citep{str2003,rag2009}. Both components of \sigcrb\ are young stars, which are on or close to zero main sequence. The components are very similar in mass, stellar radius, effective temperature and evolutionary status \citep{str2003}.

Since both of the components are rapid rotators, \sigcrb\ is an ideal target for Doppler imaging and the two stars can be imaged simultaneously. \citet{str2003} presented the first Doppler images of both stellar components of \sigcrb, and revealed dominant cool polar spots as well as equatorial warm belts. The warm belt are located at the trailing hemisphere of each star with respect to the orbital motion. \citet{don1992} detected magnetic signatures for the G-type star from the Stokes V profiles, but did not for the F-type component. The Stokes V variations are mainly located in the profile core rather than in the wings, which suggests high-latitude magnetic field on the G-type component \citep{don1992}. The surface magnetic fields of both components were detected by \citet{ros2018} with the Zeeman-Doppler imaging technique, based on the polarization spectra data collected in 2014 and 2017, and the field strength of the G0 star is significant stronger than that of the F9 star during their observing seasons.

We continued to monitor spot activity of a series of active binaries \citep{gu2003,xiang2014,xiang2015,xiang2016,xiang2020}. In this work, we present the Doppler images of both components of \sigcrb, based on the spectral data collected in 2015 March-April. We shall describe the observations and data reduction in the Section 2. The surface images will be given and discussed in Section 3 and 4, respectively. We shall summarize the final results in Section 5.

\section{Observations and data reduction}

High-resolution spectroscopic observations of \sigcrb\ were carried out with the 1m Hertzsprung SONG telescope at the observatorio del Teide, Tenerife, Spain, from 2015 March 27 to April 15. Stellar Observations Network Group (SONG) plans to construct a global network of 1m robotic telescopes, and the Hertzsprung SONG telescope is the first node \citep{and2014,and2019}. The telescope is equipped with a high-resolution \`{e}chelle spectrograph \citep{gru2017}. The resolution of the observed spectra is ~75 000 and the coverage is 4400--6900\AA. For each frame, the exposure time was set to 600s, which is corresponding to 0.6 per cent of the rotational period of \sigcrb.

A total of 361 spectra were collected during the observing run. The 1D spectra were extracted with the SONG pipeline \citep{cor2012,ant2013}. \sigcrb\ has a nearly integral-day period, which makes it difficult to observe effectively. We have observations of 11 nights, which are sufficient for two independent Doppler images. Thus these observed spectra were split into two data sets. One contains 179 spectra spanning from 2015 March 27 to April 7, and the other one contains 182 spectra spanning from 2015 April 8 to April 15. Both data sets provide a very good phase coverage and dense sampling for the Doppler imaging of \sigcrb. The largest phase gap in the first data set is 0.04, and in the second data set is 0.05. We summarize the observations of \sigcrb\ of each night in 2015 in Table \ref{tab:log}.

\begin{deluxetable}{lccc}
\tabletypesize{\scriptsize}
\tablecolumns{4}
\tablewidth{0pt}
\tablecaption{A summary of our observations on \sigcrb\ in 2015.}
 \label{tab:log}
\tablehead{
 \colhead{UT Start}&
 \colhead{UT End}&
 \colhead{No. of frames}&
 \colhead{Exp. time}\\
}
\startdata
  03-27 23:49 & 03-28 06:18 & 39 & 600\\
  03-31 23:02 & 04-01 06:13 & 39 & 600\\
  04-02 22:57 & 04-03 06:17 & 44 & 600\\
  04-04 22:35 & 04-05 03:10 & 28 & 600\\
  04-06 23:58 & 04-07 04:48 & 29 & 600\\
  04-08 22:19 & 04-09 06:10 & 47 & 600\\
  04-10 22:41 & 04-11 06:00 & 44 & 600\\
  04-11 22:07 & 04-12 06:05 & 12 & 600\\
  04-12 23:22 & 04-13 06:02 & 33 & 600\\
  04-14 00:13 & 04-14 00:33 & 3  & 600\\
  04-14 22:46 & 04-15 05:55 & 43 & 600\\
\enddata
\end{deluxetable}

We applied the Least-Squares Deconvolution (LSD; \citealt{don1997}) to combine all available atomic lines to obtain an average line profile with much higher signal-to-noise ratio (SNR). The stellar lines were extracted from the Vienna Atomic Line Database (VALD; \citealt{kup1999}) for a model atmosphere of $T_{\mathrm{eff}}$ = 6000K and log g = 4.5. The values of $T_{\mathrm{eff}}$ and log g of \sigcrb\ are taken from \citet{str2003}. The central wavelength and depth of spectral lines were used to create a line list which is required by the LSD computation. The lines within the regions of strong chromospheric (e.g. Na Double, H$_{\alpha}$) and telluric lines were removed from the list. We also derived the deconvolved telluric line profiles from the observed spectra to estimate the instrument shifts in the wavelength calibration and corrected them \citep{cam1999}.

\begin{figure}
\centering
\includegraphics[width=0.5\textwidth]{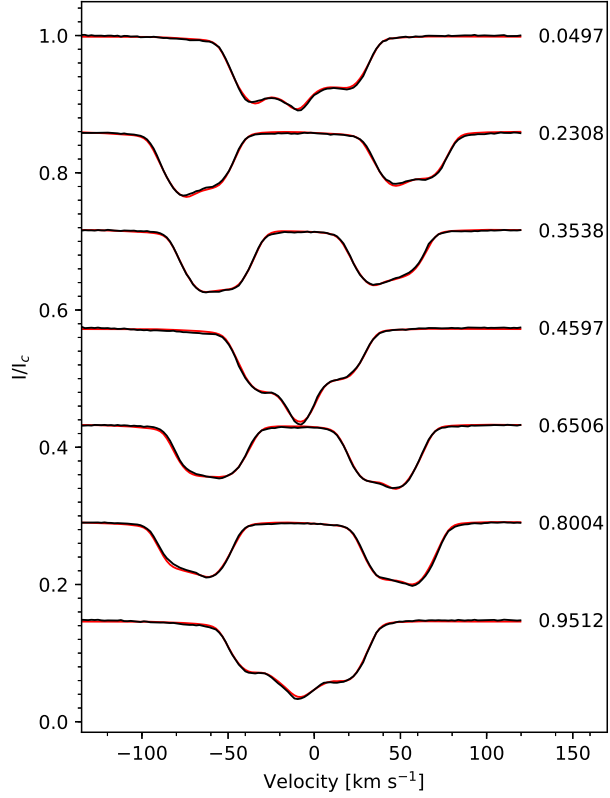}
\caption{Examples of the observed (black) and the modeled (red) LSD profiles of \sigcrb. The rotational phase is annotated beside each profile.}
\label{fig:example}
\end{figure}

\begin{figure}
\centering
\includegraphics[width=0.45\textwidth]{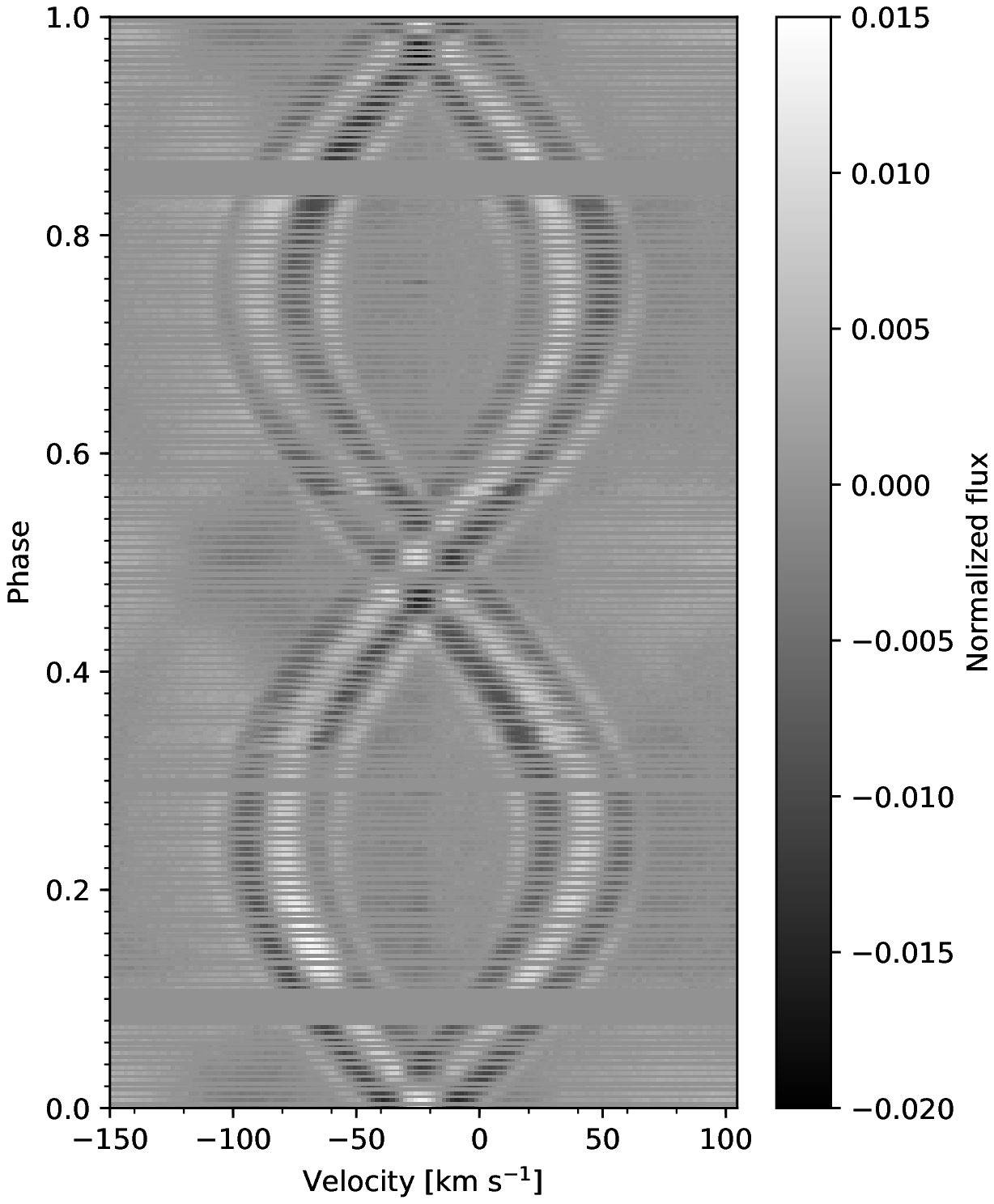}
\includegraphics[width=0.45\textwidth]{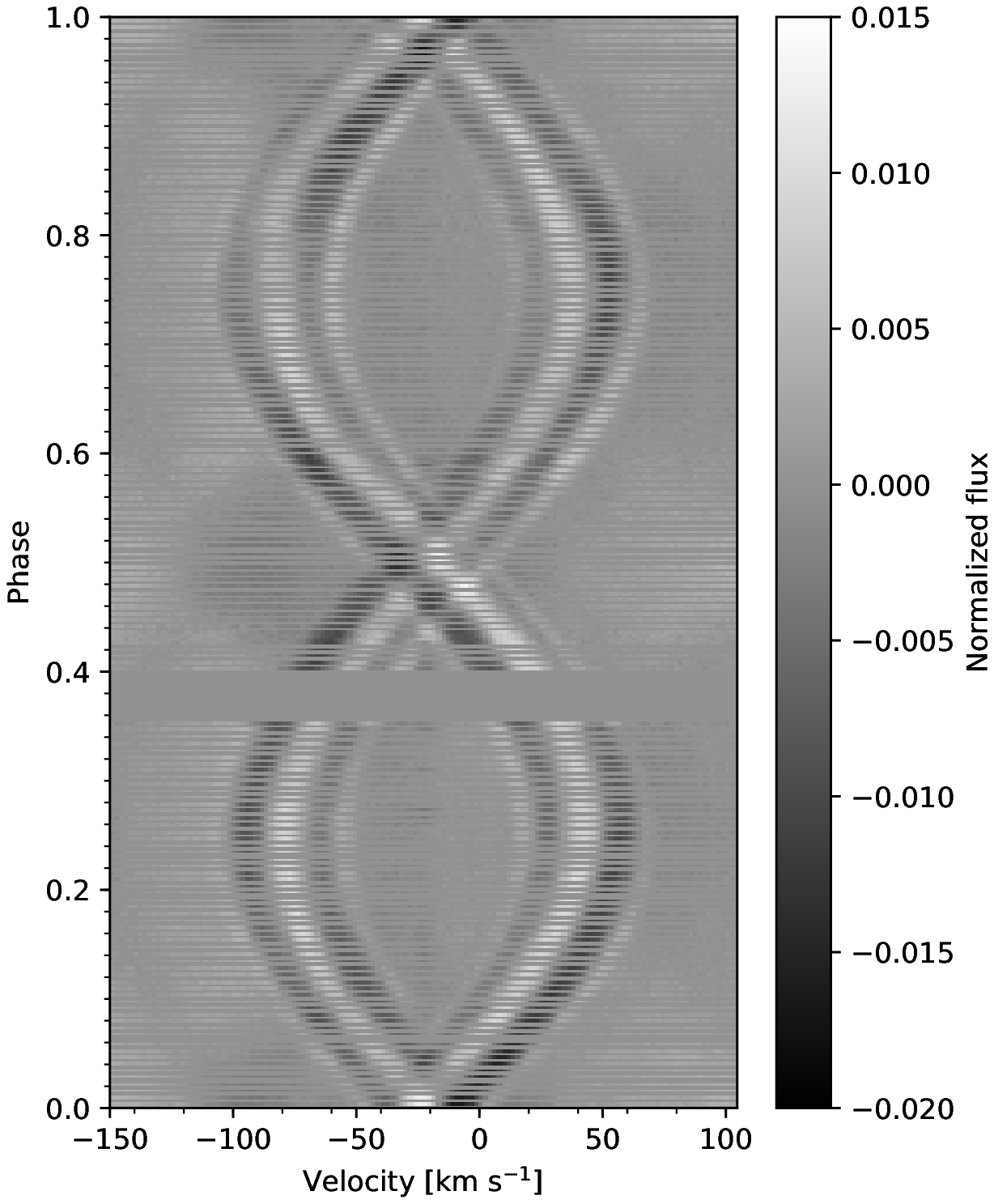}
\caption{Phased time-series LSD profiles of the two data sets after subtraction of the unspotted profiles.}
\label{fig:profile}
\end{figure}

We show examples of the resulting LSD profiles of \sigcrb\ and the corresponding modeled profiles in Fig. \ref{fig:example}. The line profiles of both components exhibit significant distortions, which indicates the presence of starspots on the two stars. We also show the phased time-series LSD profiles of two data sets in gray-scale in Fig. \ref{fig:profile}. In order to show spot signatures more clearly, an unspotted profile was subtracted from each observed profile. The time-series LSD profiles of the two components were dominated by strong signatures around the line center, which implies pronounced polar spots on the surfaces of both stars of \sigcrb.

Three inactive, slowly-rotating template stars, HR 4540 (F9V), HR 4277 (G1V) and HR 5616 (K2III), were also observed with the same instrument setup during the observing run. Since our Doppler imaging code employs the two-temperature model, which treats the surface of a star as a combination of two components, hot photosphere and cool spot, it requires a pre-calculated lookup table containing the local intensity of each component at each limb angle. Thus the spectra of template stars were deconvolved in the same manner to mimic the photospheres and spots of two stars. We used linear interpolation of the limb-darkening coefficients derived by \citet{cla2012} and \citet{cla2013} for UBVRI passbands to obtain the values at the centroidal wavelength for the photosphere and spot temperatures of each component star, and 30 limb angles were used for producing the lookup tables.

\section{Doppler imaging}

\subsection{Stellar parameters}

A reliable stellar surface reconstruction demands accurate system parameters, and errors in these parameters lead to spurious features \citep{cam1994}. The parameters of \sigcrb\ have been well determined by \citet{str2003} and \citet{ros2018}. Thus, in our case, we adopted the refined orbital elements, including the mass ratio (q), the radial velocity amplitudes of two stars (K), the inclination degree (i), the conjunction time (T$_{0}$), and the orbital period, derived by \citet{ros2018}, and list them in Table \ref{tab:par}. We fine-tuned the radial velocity of the mass center of the binary system ($\gamma$) with the $\chi^{2}$ minimization method \citep{bar2000}, and the result is also shown in Table \ref{tab:par}. Note that this value is the zero point for our data set and not the true value, since we cannot determine the systematic instrument shift in our data sets without the radial velocity standard. We also tried to determine the projected equator rotational speed (\vsini) through the same method. The result of both components is \vsini\ = 26 km s$^{-1}$, which is well in accordance with that derived by \citet{str2003}.

\begin{deluxetable}{lcc}
\tabletypesize{\scriptsize}
\tablecolumns{3}
\tablewidth{0pt}
\tablecaption{Adopted stellar parameters of \sigcrb\ for Doppler imaging. The F0 component is defined as the primary and the G0 star is the secondary.}
 \label{tab:par}
\tablehead{
 \colhead{Parameter}& 
 \colhead{Value}&
 \colhead{Ref.}\\
}
\startdata
 $q=M_{2}/M_{1}$ & 0.9724 & a\\
 $K_{1}$ (km s$^{-1}$) & 61.366 & a\\
 $K_{2}$ (km s$^{-1}$) & 63.106  & a\\
$i$ (\degr) & 28 & a\\
$\gamma$ (km s$^{-1}$) & -12.4 & DoTS\\
 T$_{0}$ (HJD) & 2450127.9054 & a\\
 Period (d)  & 1.13979045 & a\\
 \vsini~$_{1}$ (km s$^{-1}$) & $26$ & b\\
 \vsini~$_{2}$ (km s$^{-1}$) & $26$ & b\\
\enddata
\tablecomments{References: a. \citet{ros2018}; b. \citet{str2003}.}
\end{deluxetable}

\subsection{Spot images}

\begin{figure}
\centering
\includegraphics[width=0.45\textwidth]{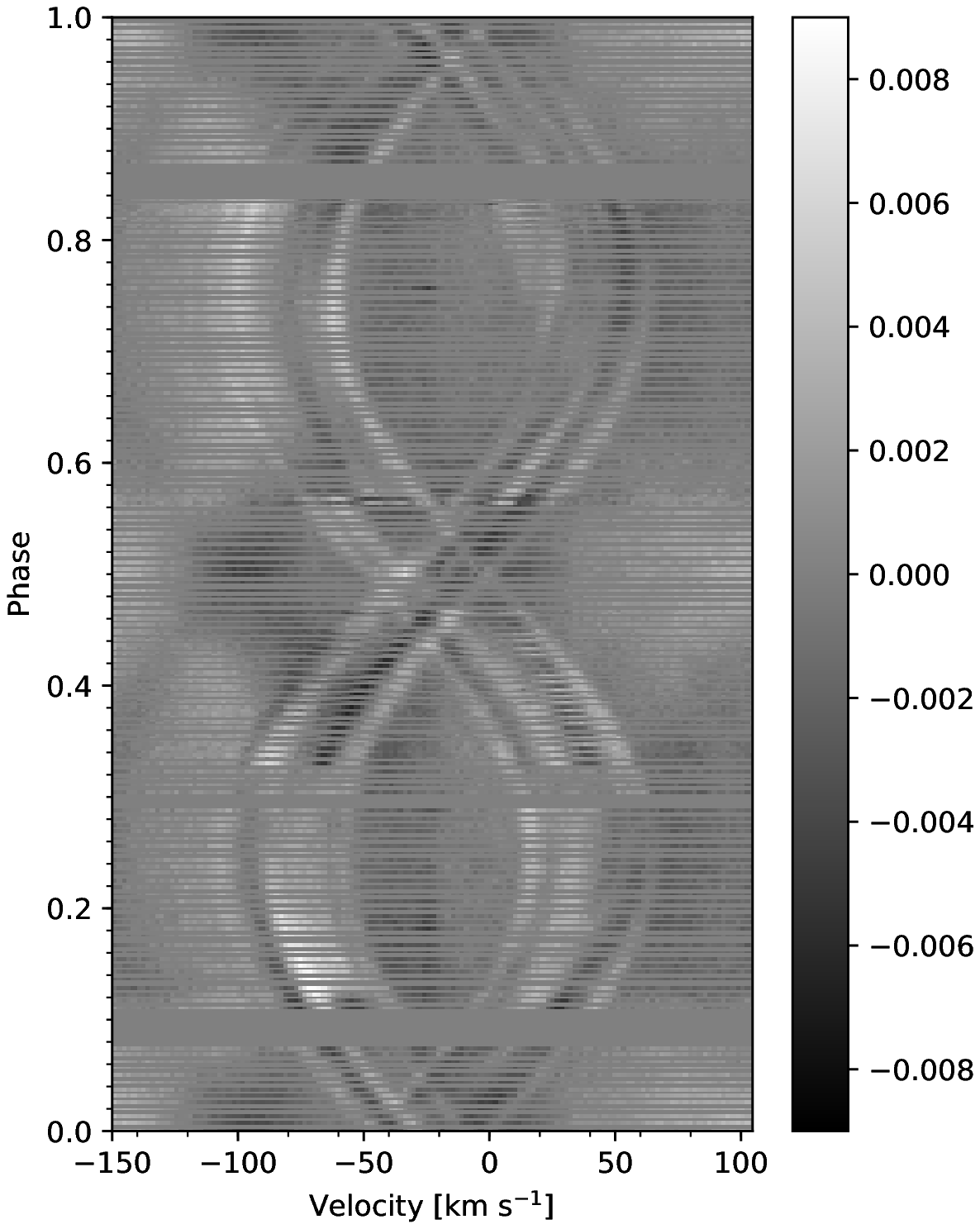}
\includegraphics[width=0.45\textwidth]{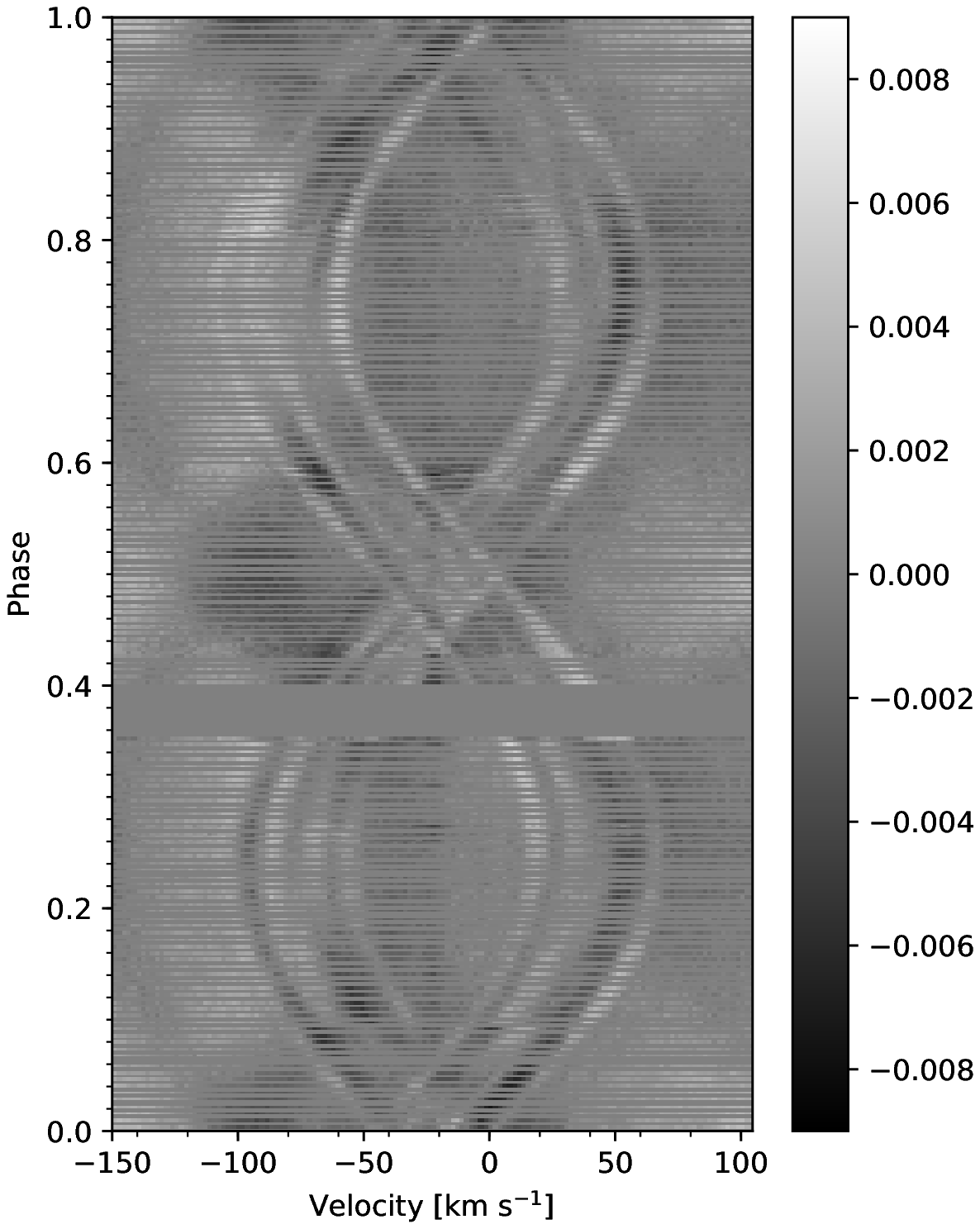}
\caption{Residuals between the observed and modeled LSD profiles. White color means the observed data point is above the fit.}
\label{fig:residual}
\end{figure}

\begin{figure}
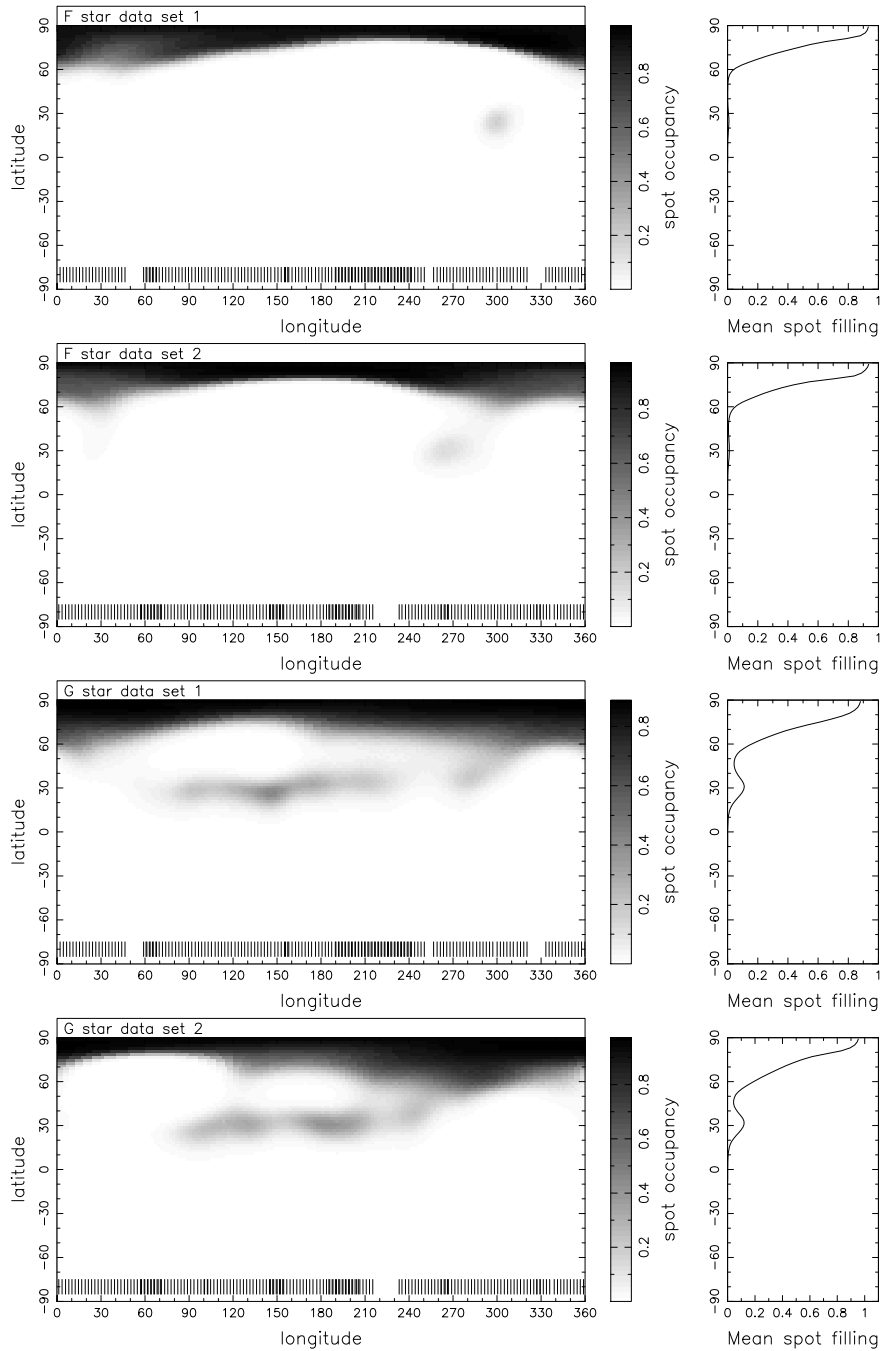

\centering
\includegraphics[bb = 72 18 360 776, angle=270, width=0.65\textwidth]{ifd1.eps}
\includegraphics[bb = 72 18 360 776, angle=270, width=0.65\textwidth]{ifd2.eps}
\includegraphics[bb = 72 18 360 776, angle=270, width=0.65\textwidth]{igd1.eps}
\includegraphics[bb = 72 18 360 776, angle=270, width=0.65\textwidth]{igd2.eps}
\caption{Doppler images of two stars. The observed phases are marked by the vertical ticks. The mean spot filling factor vs latitude is plotted beside each image.}
\label{fig:image}
\end{figure}

\begin{figure}
\centering
\includegraphics[bb = 0 0 438 479, angle=270, width=0.23\textwidth]{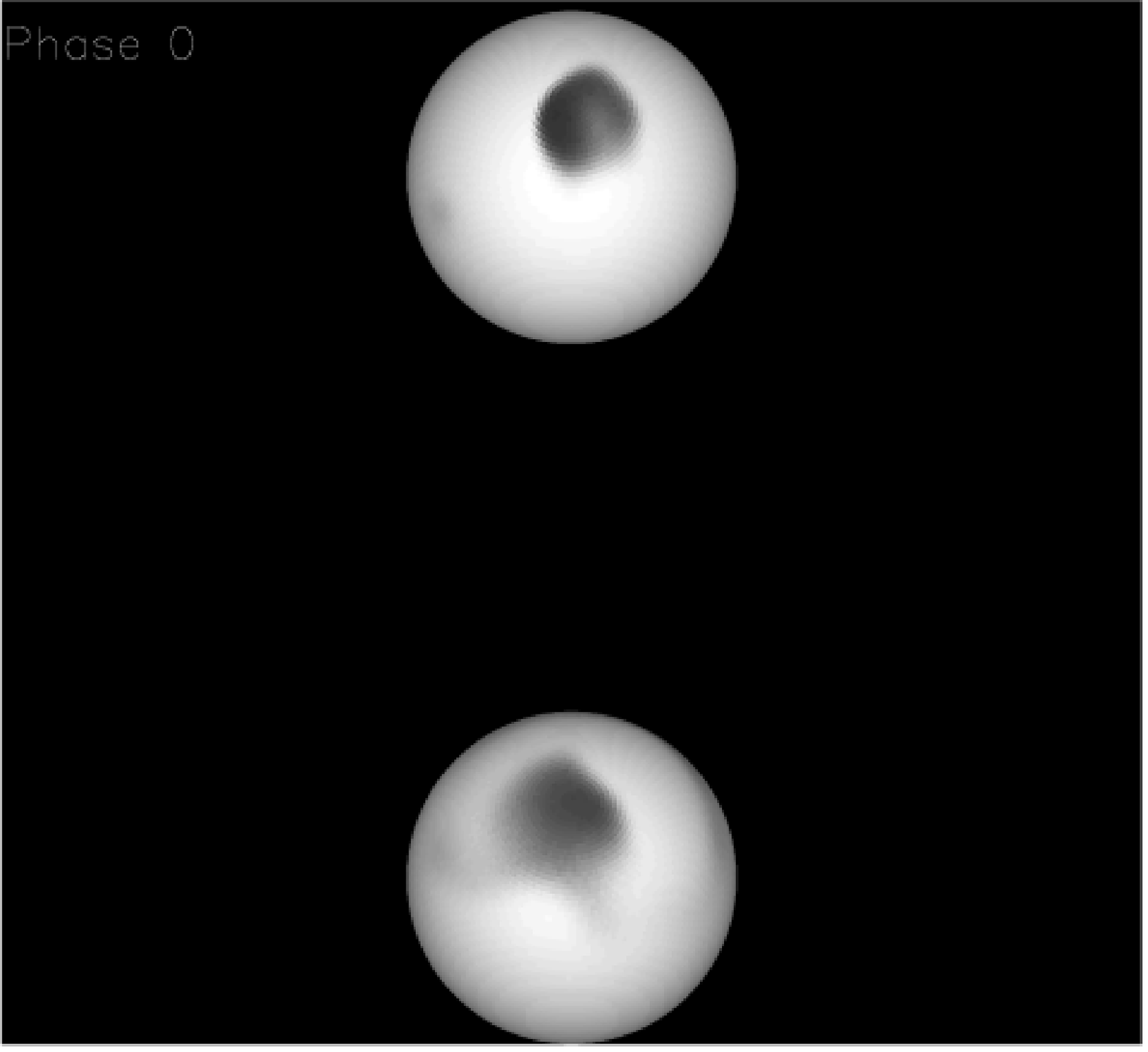}
\includegraphics[bb = 0 0 438 479, angle=270, width=0.23\textwidth]{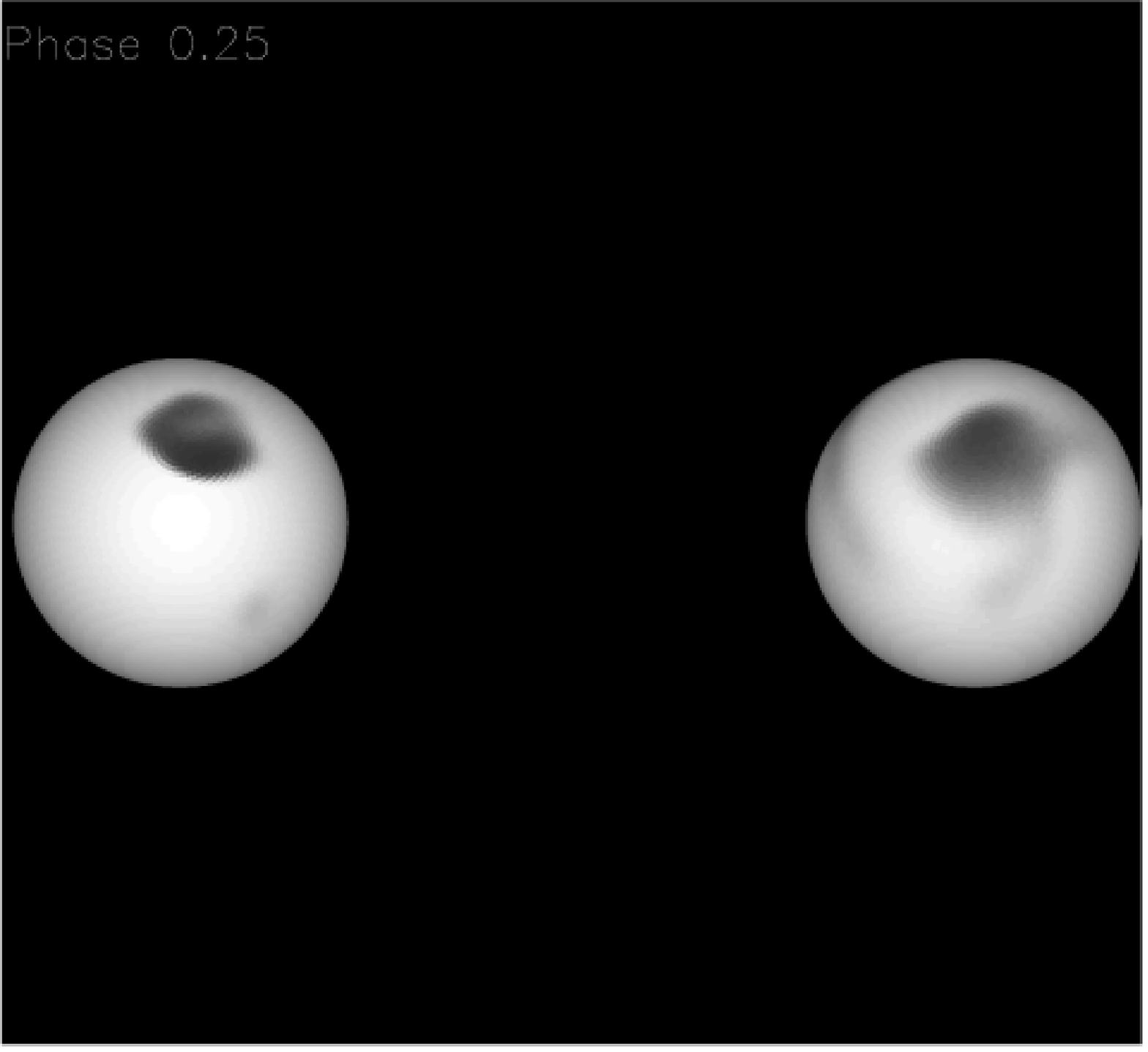}
\includegraphics[bb = 0 0 438 479, angle=270, width=0.23\textwidth]{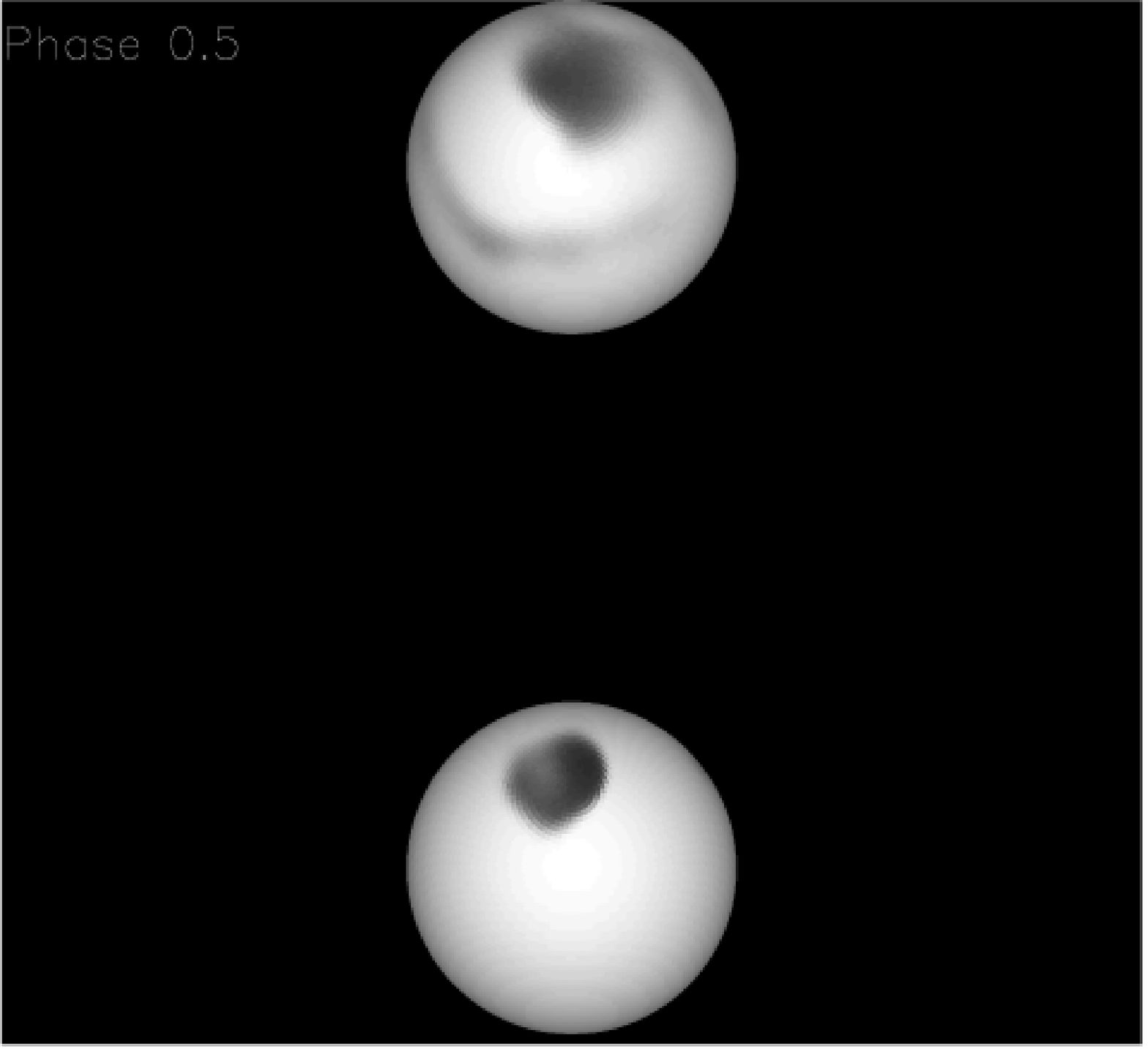}
\includegraphics[bb = 0 0 438 479, angle=270, width=0.23\textwidth]{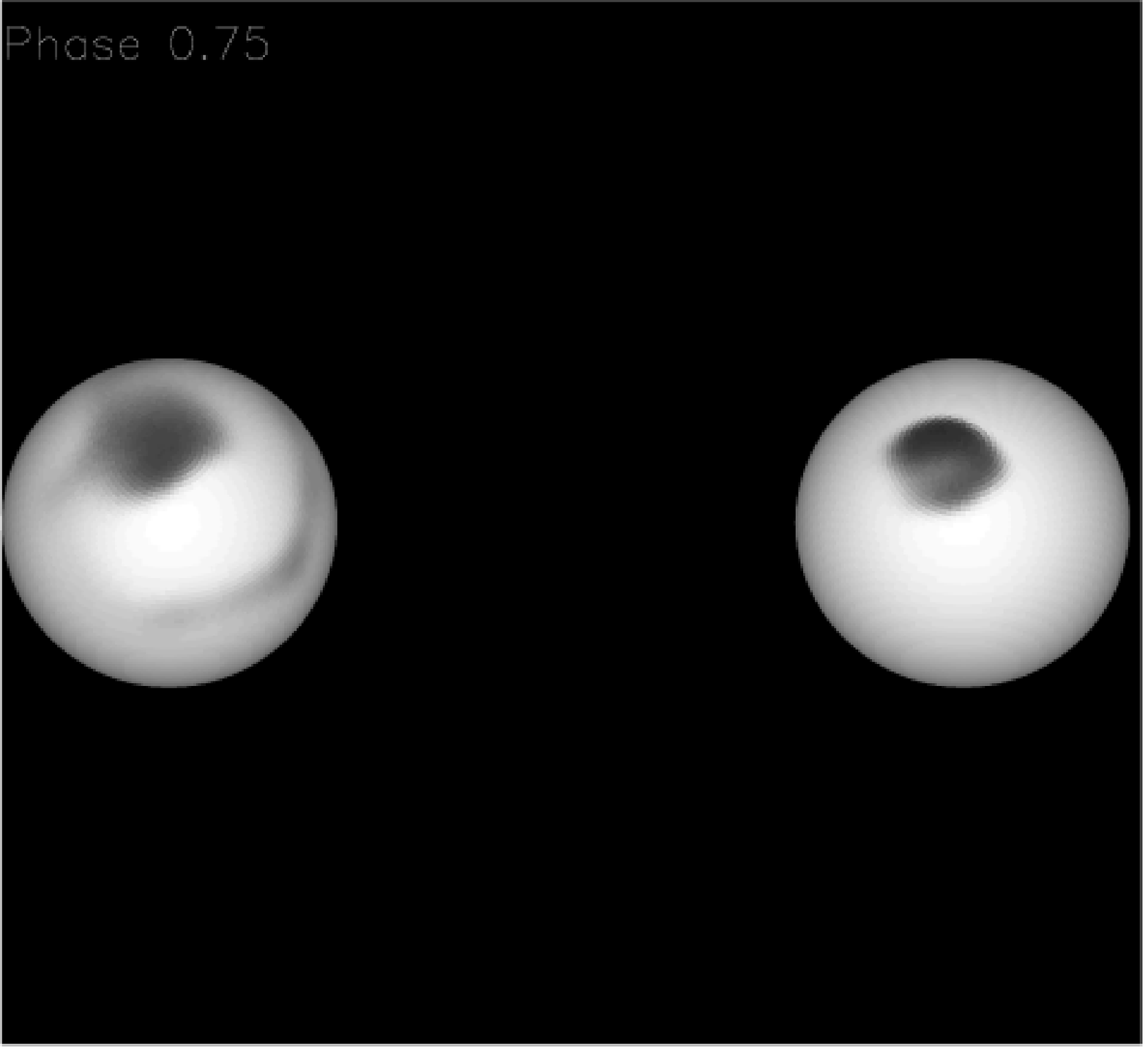}
\caption{Images of the two components of \sigcrb, at phases 0, 0.25, 0.5, 0.75, for data set No. 1.}
\label{fig:d1}
\end{figure}

\begin{figure}
\centering
\includegraphics[bb = 0 0 438 479, angle=270, width=0.23\textwidth]{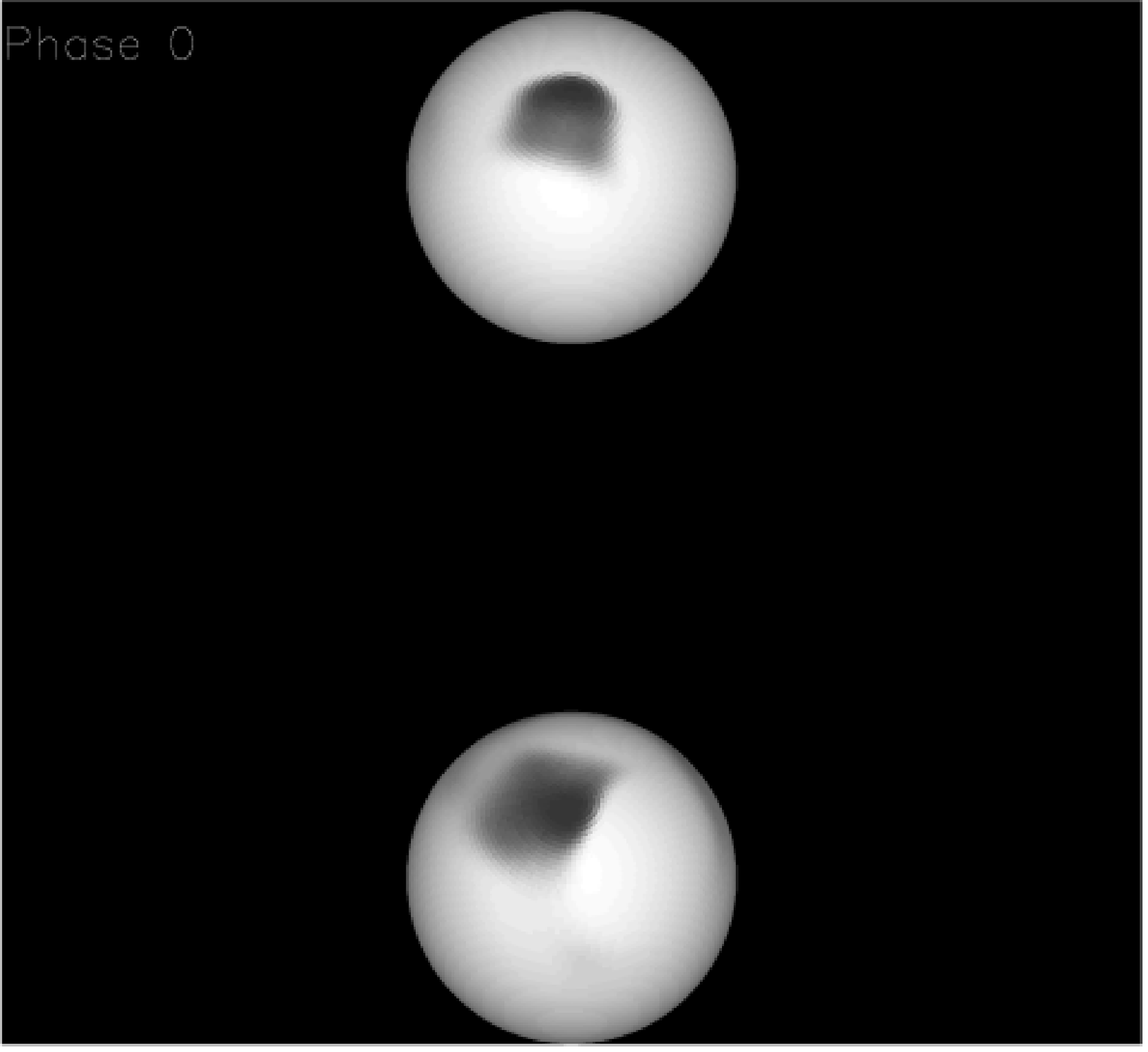}
\includegraphics[bb = 0 0 438 479, angle=270, width=0.23\textwidth]{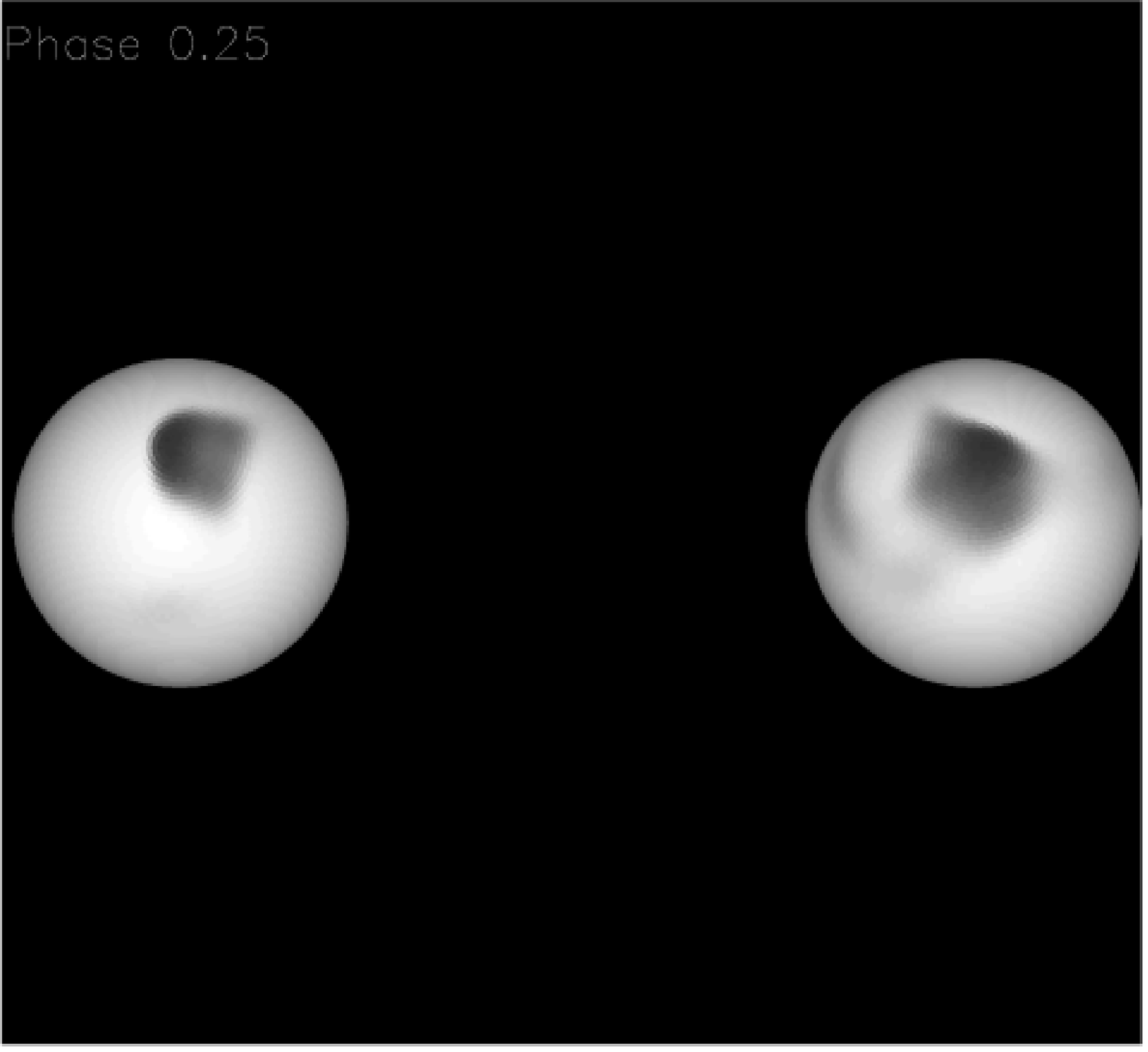}
\includegraphics[bb = 0 0 438 479, angle=270, width=0.23\textwidth]{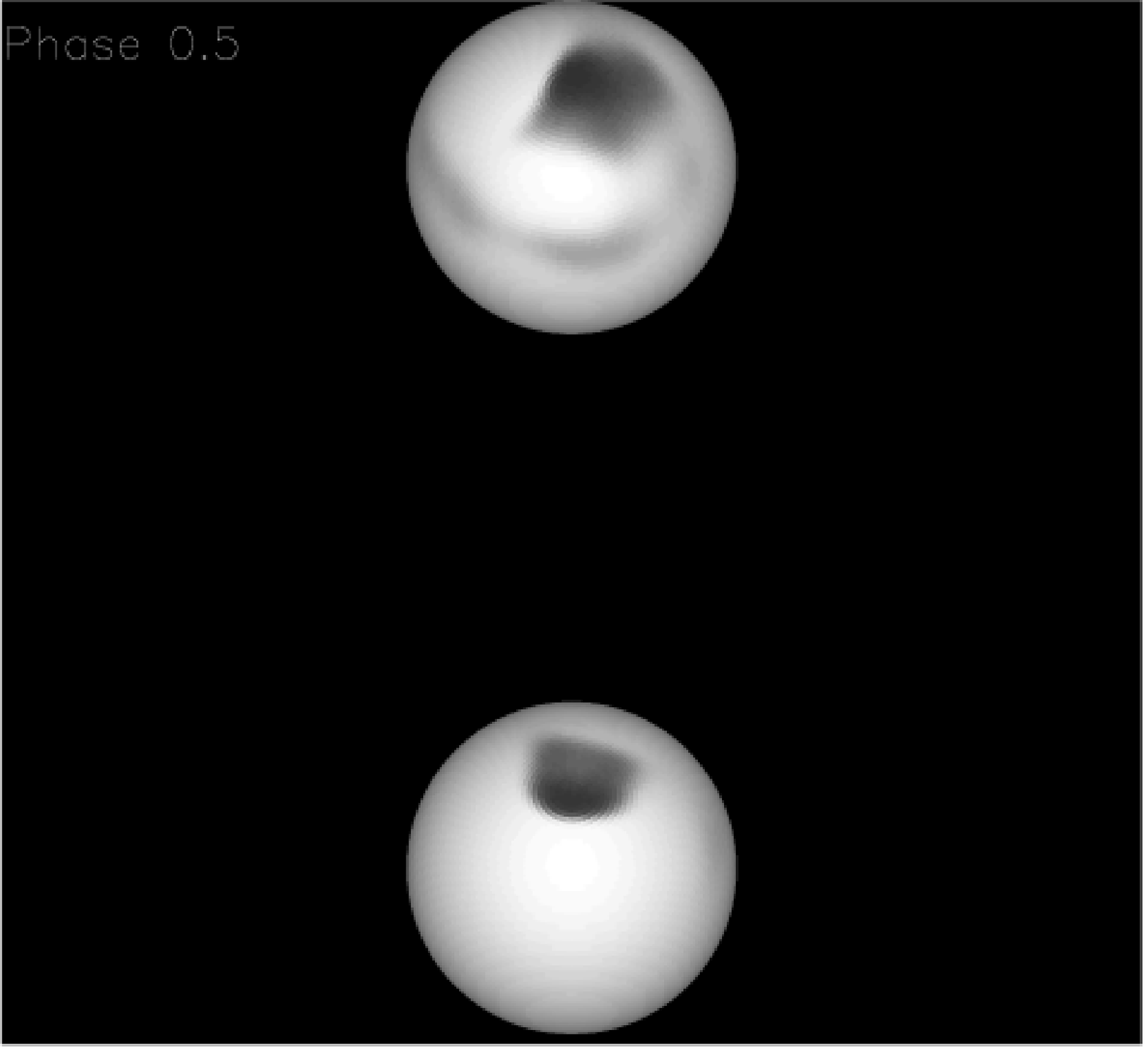}
\includegraphics[bb = 0 0 438 479, angle=270, width=0.23\textwidth]{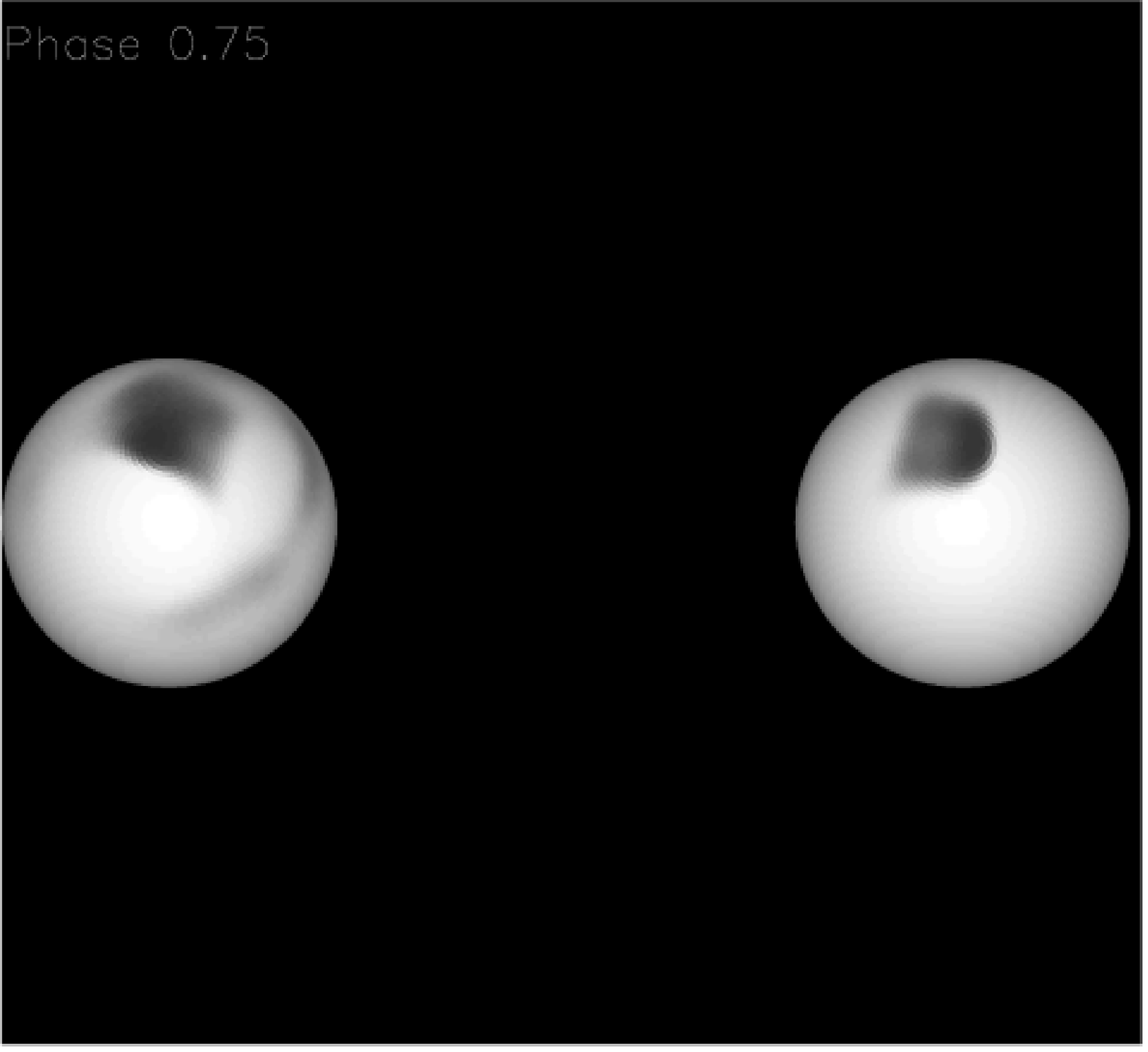}
\caption{Same as Fig. \ref{fig:d1}, but for data set No. 2.}
\label{fig:d2}
\end{figure}

We use the Doppler imaging code DoTS (short for DOppler Tomography of Stars) developed by \citet{cam1992, cam1997} to perform the maximum entropy iterations to both data sets. The residuals between the observed LSD and modeled profiles were displayed in grey scale in Fig. \ref{fig:residual}. Fig. \ref{fig:image} shows the reconstructed surface images of the two components of \sigcrb. The mean spot filling factor as a function of latitude is plotted beside each image. Note that longitude 0\degr\ on the F9 star faces longitude 180\degr\ on the G0 star in our Doppler images. To show the relationship of the spot position on the surfaces of the two components of \sigcrb\ more clearly, we display 3D images of both components of \sigcrb\ at key orbital phases 0, 0.25, 0.5, 0.75 for two data sets in Fig. \ref{fig:d1} and Fig. \ref{fig:d2}, respectively.

Our new Doppler images show a relatively simple spot pattern on the F-type component of \sigcrb. The main feature of the F9 star is a pronounced polar spot, which is consistent with what we see in the time-series LSD profiles (Fig. \ref{fig:profile}). A very weak spot feature is present at both images of F9 star, but its longitude was changed from 300\degr\ to 270\degr\ during our observations. The surface of the G-type component of \sigcrb\ is also dominated by a polar spot, and it also shows an intermediate-latitude, extended spot structure between longitudes 30 and 210, connecting to the polar cap. The polar spot on each component is asymmetric with respect to the rotational axis, which is in good agreement with the results of \citet{str2003} and \citet{ros2018}.

\begin{figure}
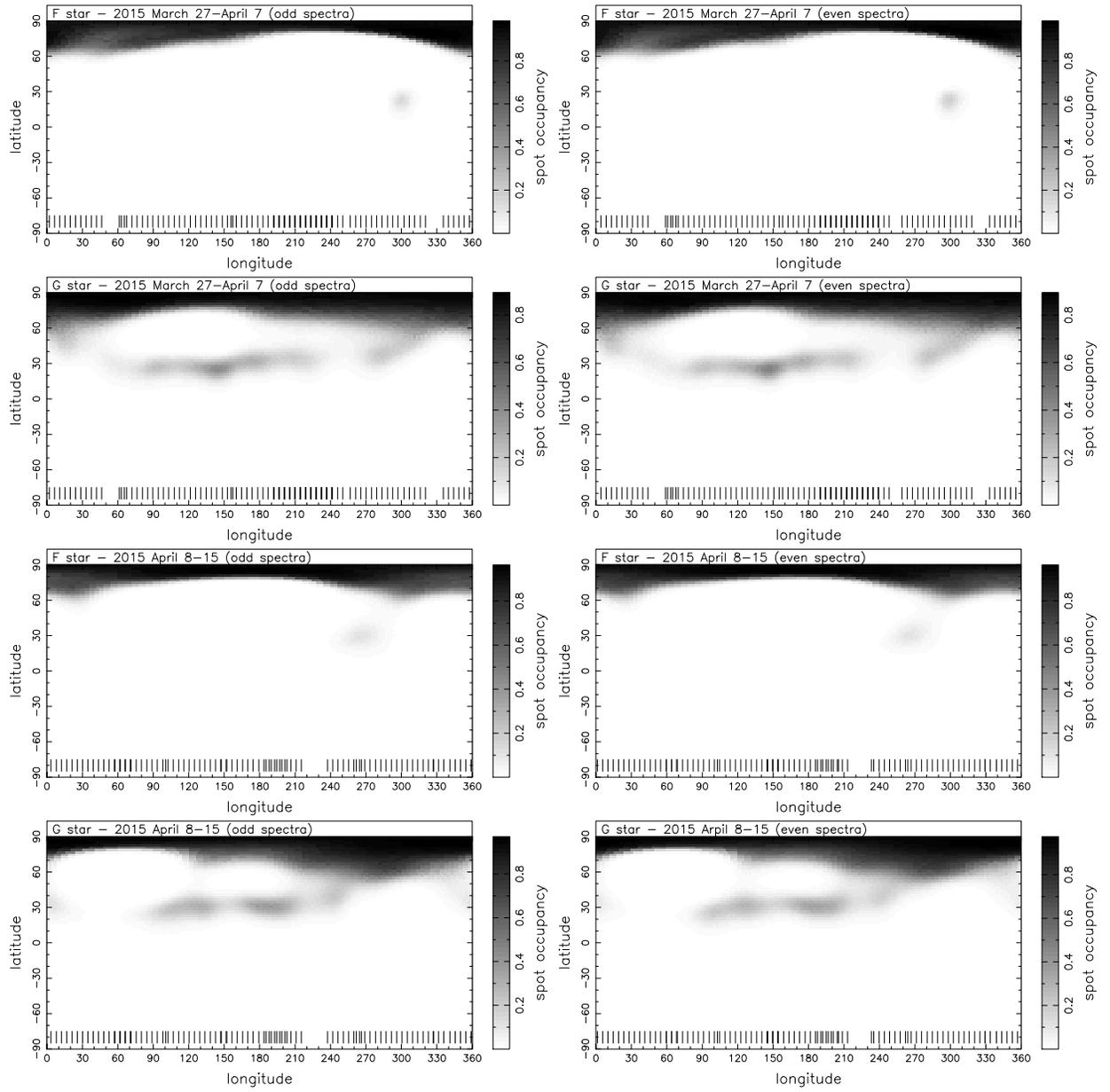

\centering
\includegraphics[bb = 72 18 360 596, angle=270, width=0.45\textwidth]{ifd1o.eps}
\includegraphics[bb = 72 18 360 596, angle=270, width=0.45\textwidth]{ifd1e.eps}
\includegraphics[bb = 72 18 360 596, angle=270, width=0.45\textwidth]{igd1o.eps}
\includegraphics[bb = 72 18 360 596, angle=270, width=0.45\textwidth]{igd1e.eps}
\includegraphics[bb = 72 18 360 596, angle=270, width=0.45\textwidth]{ifd2o.eps}
\includegraphics[bb = 72 18 360 596, angle=270, width=0.45\textwidth]{ifd2e.eps}
\includegraphics[bb = 72 18 360 596, angle=270, width=0.45\textwidth]{igd2o.eps}
\includegraphics[bb = 72 18 360 596, angle=270, width=0.45\textwidth]{igd2e.eps}
\caption{Doppler images of the two components derived from the odd-numbered and even-numbered spectra of two data sets}
\label{fig:oe}
\end{figure}

In order to determine the reliability of the surface features in our new Doppler images of \sigcrb, we also performed an odd-even test on the reconstructions. We split each data set into two subsets which respectively consist of odd-numbered and even-numbered LSD profiles. Then we derived the surface images of the two components from the independent subsets. As shown in Fig. \ref{fig:oe}, the reconstructed spot patterns in each image pair are nearly identical for each star, which demonstrates that our image reconstructions are reasonably reliable and that the dense sampling is helpful for the Doppler imaging.

\subsection{Differential rotation}

The Doppler imaging can spatially resolve the stellar disc and thus can be used to measure the surface differential rotation. One can obtain the surface shear rate with either the cross-correlation \citep{don1997r} or the image shear method \citep{pet2002}. The cross-correlation method requires observations spanning at least two rotational cycles and good phase coverage. The advantage of this method is that it does not need any prior knowledge of the differential rotation law.

Two Doppler images several rotational period apart may show latitude-dependent rotation pattern. Actually, some clues can be seen directly from our Doppler images. The appendages of polar spots moved to less longitude between the two observation epochs, indicating a slower rotation of the high-latitude feature with respect to the co-rotating frame.

\begin{figure}
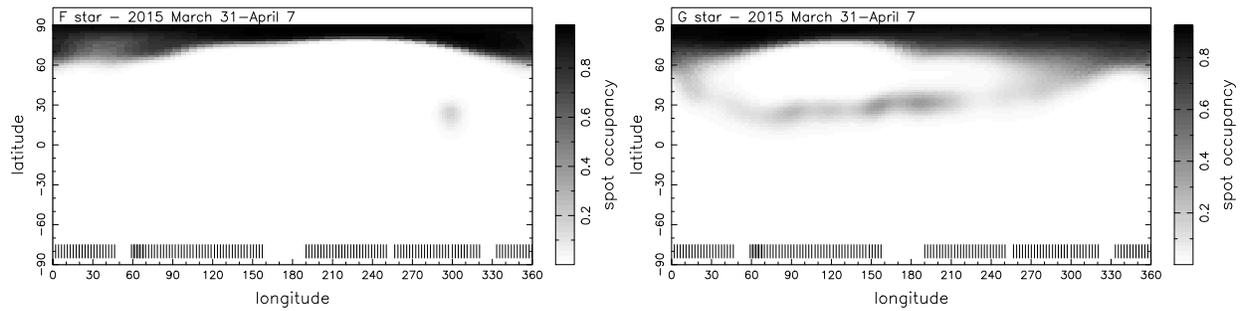

\centering
\includegraphics[bb = 72 18 360 596, angle=270, width=0.45\textwidth]{if1.eps}
\includegraphics[bb = 72 18 360 596, angle=270, width=0.45\textwidth]{ig1.eps}
\caption{Doppler images of both components derived from data set No. 1 without observations in the night of 2015 March 27.}
\label{fig:newimages}
\end{figure}

\begin{figure}
\centering
\includegraphics[width=0.45\textwidth]{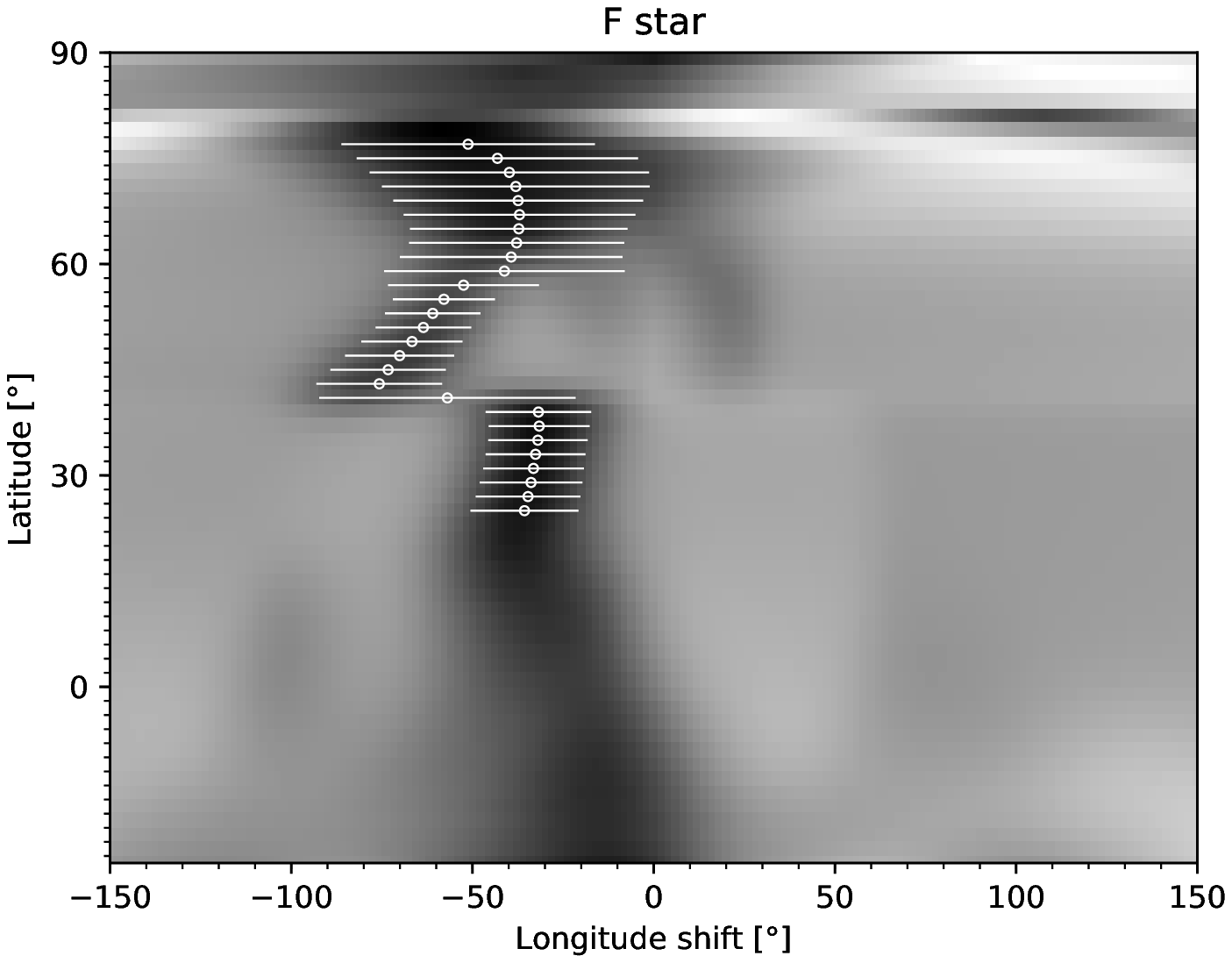}
\includegraphics[width=0.45\textwidth]{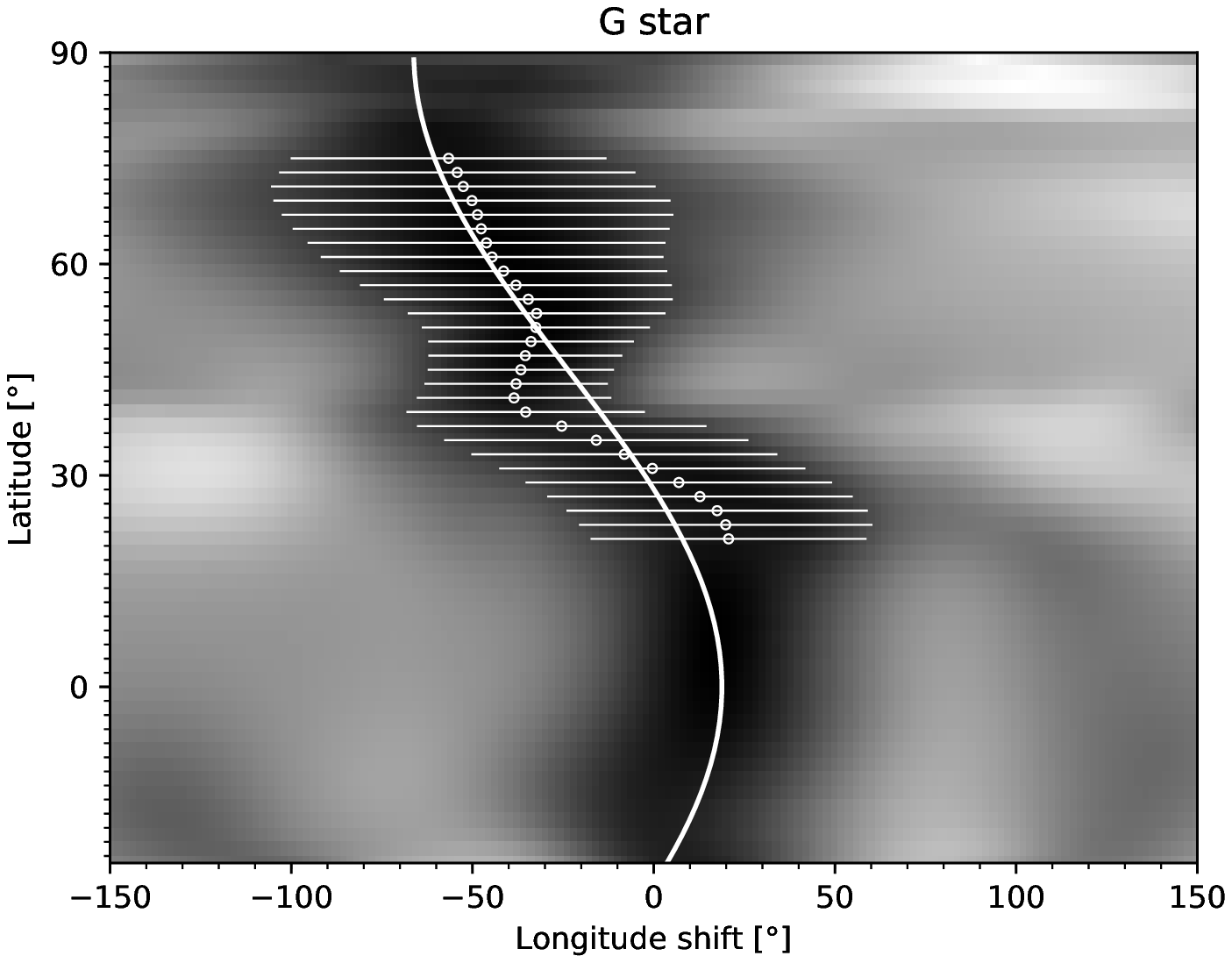}
\caption{Cross-correlation maps of each latitude for the two components. The peak (white open circle) of the cross-correlation function for each latitude belt is determined by fitting a Gaussian profile and the full width at half maximum (FWHM) of the Gauss profile is presented by the white horizontal line. The white curve is a solar-like differential rotation fit.}
\label{fig:dr}
\end{figure}

The first 39 spectra of data set No. 1 are collected on 2015 March 27, when is 4 days before the second observing night of 2015 March 31. This time interval is much larger than the rest and thus may introduce more errors in the estimate of the differential rotation. Hence, for the cross-correlation study, we firstly derived the Doppler images of the two components from data set No. 1 without the spectra of March 27. The new surface images are shown in Fig. \ref{fig:newimages} and the phase coverage is still sufficient. Then we used these images and the ones derived from data set No. 2 to calculate the cross-correlation function of each latitude stripe on each component (Fig. \ref{fig:dr}). The peak of the cross-correlation function of each latitude was determined by fitting a Gaussian profile. We tried to fit the peaks with a solar-like surface shear curve as follows
\begin{equation} \Omega (l) = \Omega_{eq} - \Delta \Omega \sin^{2} l,\end{equation}
where $l$ is the latitude and $\Omega_{eq}$ is the rotational speed at equator and $\Delta \Omega$ is the difference between the equator and the stellar pole. The points between latitudes 20\degr\ and 70\degr\ were used, because the Doppler imaging technique has a poor longitude resolution at higher latitudes and is insensitive to the latitude of spots near the stellar equator, and latitudes below 20\degr\ were relatively featureless. The G0 component of \sigcrb\ shows a clear solar-like differential rotation, and the surface shear rate is $\Delta \Omega = 0.180 \pm 0.004$ rad d$^{-1}$ and $\alpha = \Delta \Omega / \Omega_{eq} = 0.032 \pm 0.001$. However, we failed to fit the cross-correlation map of the F9 component. It is probably due to that most spots on the F9 star are concentrated at high-latitude and no significant spot is present at mid-to-low latitudes.

\section{Discussion}

We have presented new Doppler images of both components of the double-lined binary \sigcrb, based on two data sets with very good phase coverage collected in 2015 March and April. The maximum entropy reconstructed images show that the surfaces of both components are dominated by pronounced polar spots. On each star, the polar spot is asymmetric with respect to the stellar pole. At latitude 30\degr, the G0 star also exhibits an extended spot structure connecting to the polar spot whereas the F9 star only shows a weak feature.

The epoch of our observations is just between two observing seasons of \citet{ros2018}. The spot patterns in our surface images are compatible to their reconstructed brightness distributions, especially for high latitude features. The polar spots on two components are revealed by all Doppler images. This may indicate a long lifetime for the polar spots on two stars, at least three years, which is common on active binaries (e.g. \citealt{str2009}). The spot configurations of mid-to-low latitudes in our images are different to those in their images, which should be due to the spot evolution.

Our Doppler images of the two components of \sigcrb\ are very compatible to those of \citet{str2003}. They revealed the coexistence of dominant polar cool spots on both stars and the presence of a mid-to-high latitude cool spot on the F9 star. They also found equatorial warm belt on the trailing hemisphere of each star with respect to the orbital motion. Since our Doppler imaging code uses the two-temperature model, we cannot reconstruct any hot spot on the stellar surface. The spot structure at latitude 30\degr\ on the G0 star is facing the F9 star, as seen in Fig. \ref{fig:d1} and \ref{fig:d2}. This is somewhat similar to the spot distribution of the two components of ER Vul \citep{xiang2015}. However, considering the images of \citet{ros2018} in 2014 and 2017, the spot longitude distributions of both stars seem to be chaotic and they do not show any clear preferred active longitudes, which are found on many active binary and single stars \citep{str2009}. Apparently, any conclusive result on that requires more reproducible observations.

The spot coverages of the northern hemispheres of the F9 and G0 components of \sigcrb, derived from the reconstructed surface images, are 4.5\% and 10.5\%, respectively. The G0 star showed a higher level of spot activity than the F9 star in 2015. This difference is consistent with the results of \citet{don1992} and \citet{ros2018}. \citet{don1992} only detected magnetic signatures on the G0 star but not on the F9 star, based on the polarisation spectra data collected in 1990 and 1991. \cite{ros2018} modeled the Stokes V profiles of \sigcrb\ and detected magnetic field of both components, but the field strength on the G0 star is larger than that on the F9 star.

Considering the similarity of the two components of \sigcrb, it is interesting to see this persistent difference of magnetic activity levels of two stars in various observation seasons. The F9 and G0 stars have very close effective temperature, stellar mass, radius and evolutionary status \citep{str2003}, which means that they should have similar depth of the convection envelope. The rotational periods of the two stars are equal to each other due to the tidal lock. These are believed to be the most important factors of the stellar dynamo, which generates magnetic field on stars. \citet{ros2018} inferred that the significant difference in magnetic field strengths of the two stars is not likely caused by their long-term activity cycle, because the same difference was also revealed by \citet{don1992} in 1990--1991 and it seems to be persistent for 30 years, and that the strength difference is probably related to the different regimes of the stellar dynamo with low Rossby numbers, which can produce either a strong, dipole magnetic field or a weak, complex field on stars as revealed by \citet{gas2013}.

The two solar-like components (G0V + G2V) of ER Vul are very similar to those of \sigcrb. The rotational period of ER Vul is 0.7d, shorter than that of \sigcrb. \citet{pis2001} presented the Doppler images of both components of ER Vul and found large temperature variations and the presence of hot spots at sub-stellar points on the two stars. In our previous work \citep{xiang2015}, we also derived Doppler images of ER Vul for 2006 and 2008 observing seasons. The spot patterns are more complex than those on \sigcrb. Both components of ER Vul show spots at various latitudes, from the equator to the stellar pole, and most mid-to-low latitude spots are concentrated at the hemisphere facing another star.

Another similar star is AF Lep, which is a single star with a similar spectral type (F8/9) and rotational period (1.0 d). The surface images derived by \citet{jar2015} indicate the presence of a dominant high-latitude spot, similar to that on F9 star of \sigcrb. Their theoretic models suggested that the radiative interior and the convection zone at the equator should have a same rotational speed to produce a high-latitude magnetic field.

A difference between the photometric and orbital periods was found by \citet{str2003} and \citet{ros2018}. \citet{ros2018} estimated that the rotational periods of the F9 and G0 star are 0.039 and 0.024 d longer than the orbital period. They attribute this to the surface differential rotation and the magnitude should be 0.1--0.19 rad d$^{-1}$. The larger rotational period may indicate that the dominant high-latitude features rotate more slowly than the orbital co-rotation \citep{ros2018}. The slower rotation of high-latitudes features can be directly seen in our Doppler images. Although we cannot derive a clear differential rotation law for the F9 star due to the lack of mid-to-low latitude spots, we found that the high-latitude features were about 50\degr\ backward with respect to the co-rotating frame, corresponding to 6\degr\ per day. This is consistent with the estimate of \cite{ros2018}.

Through the cross-correlation method, we derived a solar-like latitude-dependent rotation for the G0 star. The estimate is $\Delta \Omega = 0.180 \pm 0.004$ rad d$^{-1}$, which is in good agreement with the inference of \cite{str2003} and \cite{ros2018}. The results indicate that the stellar equator of the G0 star rotates faster than the pole and laps it once every 35 d. Note that error estimate is just statistical, and the spot evolution may induce more errors in the estimate of the surface shear, since our observing run spans 15 nights. Further multi-site observations are needed to study the surface differential rotation of \sigcrb.

Recently, surface differential rotation on various single and binary stars has been detected, and the relationship between the surface shear rate and stellar parameters were investigated by many authors \citep{bar2005r,cam2007,kov2017}. The surface differential rotation rate increases with the decrease of the depth of the stellar convection zone \citep{bar2005r,mar2011}. The shear rate found on the G0 component of \sigcrb\ is in good agreement with the relations between the differential rotations and the effective temperatures derived by \citet{bar2005r}.

\section{Conclusion}

We obtained spectra of the double-lined binary \sigcrb\ with the 1m Hertzsprung SONG telescope during 11 nights in March and April, 2015. The time-series LSD profiles derived from the observed spectra formed two independent, well-sampled data sets from which we derived two Doppler images of each component of \sigcrb.

Our new Doppler images show dominant polar spots on both of the F9 and G0 star components of \sigcrb, similar to those derived by \citet{str2003} and \citet{ros2018}. These polar spots are asymmetric about the stellar rotational pole. The spot images also show a weak feature at intermediate latitude on the F9 star and an extended spot structure on the G0 star. The spot coverage of the G0 star is larger than that of the F9 star, which may indicate that the G star had a higher level of the magnetic activity than the F star during our observing seasons.

Differential rotation, as an important ingredient in generating stellar magnetic field, is difficult to be quantified. The surface differential rotations of only tens of active stars have been measured by means of Doppler imaging, which can spatially resolve the stellar disk. These measurements require relatively long time-span observations covering several rotational cycles and sufficient phase sampling. Differential rotation rates of a variety of stars are needed to investigate the relations to the stellar parameters, such as the rotational periods and effective temperatures, to better understand the stellar dynamo.

The cross-correlation of latitude stripes of two independent Doppler images of the G0 star reveals a solar-like surface differential rotation. The shear rate is $\Delta \Omega = 0.180 \pm 0.004$ rad d$^{-1}$, which means the equator of the G0 star rotates faster than the stellar pole and laps it once every 35 d. The relative differential rotation rate of the G0 star is $\alpha = \Delta \Omega / \Omega_{eq} = 0.032 \pm 0.001$. The cross-correlation map for the F9 star, however, does not show a clear differential rotation law, due to the lack of mid-to-low latitude features, but its high-latitude spots rotate slower with respect to the co-rotating frame.

The near 1-d period of \sigcrb\ makes it difficult to obtain complete phase coverage at only one site. In the future, with more spectroscopy telescopes of SONG network at global sites, active stars with such integral-day rotational periods will be observed more effectively. This will greatly help to investigate their starspot activity and surface differential rotation.

\section{Acknowledgements}

This work is supported by National Natural Science Foundation of China (grant Nos. 10373023, 10773027, 11333006, 11603068, and U1531121) and Chinese Academy of Sciences project (No. KJCX2-YW-T24). Funding for the Stellar Astrophysics Centre is provided by The Danish National Research Foundation (Grant DNRF106). A special thanks goes to Antonio Pimienta and the team of operators at Observatorio del Teide who have contributed to the maintenance and running of the Hertzsprung SONG Telescope. We are very grateful to the anonymous referee for valuable comments and suggestions that significantly improved the clarity and quality of this paper. This work has made use of the VALD database, operated at Uppsala University, the Institute of Astronomy RAS in Moscow, and the University of Vienna.

\bibliographystyle{aasjournal}
\bibliography{sig2crb2015}

\end{CJK*}
\end{document}